\documentclass[11pt,twoside]{article}
\usepackage{atom2010}
\usepackage{graphicx}
\usepackage{color}
\resetcounters

\newcommand{\Fe}[5]{\mbox{$#1\,^#2{\rm #3}^{{\rm #4}}_{\rm #5}$}}
\newcommand{\Co}[5]{\mbox{$#1\,^#2{\rm #3}^{{\rm #4}}_{\rm #5}$}}

\begin{document}

\title{Role of input atomic data in spectroscopic analyses of the Sun and
metal-poor stars}
\author{Maria Bergemann$^1$
\affil{$^1$ Max-Planck Institute for Astrophysics, Karl-Schwarzschild Str. 1,
85741, Garching, Germany}}

\begin{abstract}

Analysis of high-resolution stellar spectra relies heavily upon atomic data.
These include energy levels, wavelengths, cross-sections for various types of
interactions between particles and photons, such as photoionization and
collision induced transitions. Quantum-mechanical calculations provide largest
part of these data. In this paper I describe atomic data necessary to compute
model atmospheres and line formation for the late-type stars both in the
assumption of local thermodynamic equilibrium (LTE) and in a more general case
of non-LTE. I will focus on transition metals with $21 <  Z < 29$ and discuss
whether (and where) more complete and/or accurate atomic data are necessary.

\end{abstract}

\section{Introduction}

Chemical composition of late-type (FGK) stars provides major observational
constraints to the chemical evolution of the Galaxy. These stars span a huge
range of metallicities and ages and, due to relatively cool surfaces, their
spectra are rich in lines of different atomic and molecular species that allows
to determine abundances of many chemical elements. For certain elements, lines
due to forbidden and dipole-allowed transitions, as well as molecular lines, can
be analysed simultaneously, e.g. [CI], CH, and C I.

At present, determination of element abundances usually proceeds in two steps.
First, one constructs a model of a stellar atmosphere using basic conservation
laws and certain simplifying assumptions, which are valid for a restricted range
of stellar parameters. The model gives variation of basic atmospheric
quantities, local kinetic gas temperature $T_{\rm e}$ and number density of free
electrons $N_{\rm e}$, as a function of depth. Next, the model is used to solve
radiative transfer equation for individual spectral features to compare them
with observations. In both steps, it is necessary to describe correctly
absorptive properties of the atmospheric plasma, which depend on the
distribution of atoms and ions among excitation states and ionization stages
$N_i$, and on the cross-sections for interaction of gas particles with radiation
field. In a simple case of \emph{local thermodynamic equilibrium} (LTE), the
former is computed from the Saha-Boltzmann equations for local values of $T_{\rm
e}$ and $N_{\rm e}$ at each point of the model atmosphere. However, in reality
atomic level populations $N_i$ are, in return, affected by the radiation field,
which is generally highly non-local due to scattering processes. Thus, $N_i$
also depend on physical conditions at other depth points and must be determined
from the solution of full equations of statistical equilibrium for each ion and
each of its energy levels, an approach known as non-LTE. These equations must
include the rates of all processes, which represent interaction between
particles and photons, such as radiatively-induced excitation and ionization of
atoms. The need for accurate cross-sections of these processes renders atomic
data as one of the key ingredients in stellar atmosphere studies.

Below, I will briefly describe atomic data necessary to compute model
atmospheres and line formation both in LTE and non-LTE cases. I will focus on
the late-type stars with $4000 < T_{\rm eff} < 6500$ and $3 < \log g < 5$, and
on transition metals with $21 <  Z < 29$. Abundances of these elements in old
stars provide key information for understanding explosive nucleosynthesis in
supernovae. Brief reviews of atomic data used in precision stellar spectroscopy
are also given in \citet{2008psa..conf..123J} and \citet{2009PhST..134a4004M}.

\section{Opacity in model atmospheres and spectrum synthesis}{\label{sec:opac}}

Construction of theoretical model atmospheres requires a detailed knowledge of
the monochromatic absorption coefficients over a huge frequency interval. For
the range of stellar parameters corresponding to late-type stars, the following
opacity sources are relevant. The continuous extinction is represented by the
free-free transitions in \ion{H}{i}, He$^{-}$, H$^{-}$, and H$^{-}_2$, Rayleigh
scattering for H, He, and H$_2$, and Thomson scattering on electrons. Also
included are bound-free transitions in H$^{-}$, \ion{H}{i}, \ion{He}{i},
H$^{-}_2$ and H$^{+}_2$. Very important absorbers in the UV and visual spectral
range are \ion{C}{i}, \ion{Mg}{i}, \ion{Si}{i}, \ion{Al}{i}, \ion{Fe}{i}. For
these species it is crucial to have accurate photoionization cross-sections.
\citet{2004A&A...420..289G} showed that quantum-mechanical cross-sections for Fe
I \citep[][Iron project]{1997A&AS..122..167B} lead to a much better agreement
between theoretical and observed solar fluxes in the UV and IR spectral regions.
The influence of Fe I bound-free opacity on the solar UV continuum fluxes was
also demonstrated in \citet{2001ApJ...546L..65B} using MARCS and ATLAS9 models,
although they argued that the quantum-mechanical cross-sections should be
increased by a factor of two to match the fluxes in the spectral range $3000
\ldots 4000$ \AA. These same arguments may not apply to other types of
atmospheric models.

Line opacity is dominated by bound-bound transitions in neutral and
singly-ionized atoms of Fe-peak elements, although opacity due to other atoms
(e.g. \ion{H}{i}) and molecules (e.g. H$_2$, CH, OH, MgH, TiO) has to be
included as well. Modern model atmosphere codes \citep[e.g.
MAFAGS:][]{2009A&A...503..177G} use tens of millions of lines to construct
opacity distribution functions or sample them at certain frequencies to save
computational efforts.
Recent calculations of Kurucz\footnote{http://kurucz.harvard.edu/} increased the
number of predicted radiative bound-bound transitions for several neutral atoms,
such as \ion{Fe}{i}, \ion{Ti}{i}, \ion{Cr}{i} to few millions; most of these
transitions originate from the levels with very high excitation energies.
\citet{2009A&A...503..177G} showed that including all predicted transitions has
a small effect on the mean temperature structure of the solar atmosphere: up to
$25$ K in the line formation regions. Another interesting result was a reduced
discrepancy between observed and synthetic solar fluxes in the UV.

\section{Line formation}

Calculation of synthetic profiles for comparison with observations is not
trivial even under LTE and requires atomic data for individual lines:
wavelengths, energies of lower levels of transitions, oscillator strengths,
pressure broadening parameters, hyperfine splitting and isotopic shifts. For
most of these parameters, data from laboratory measurements are usually
available, however the accuracies are not always as good as desirable.

In calculations of abundances by the most reliable and rigorous method of
spectrum synthesis, one derives the product of abundance and oscillator
strength of a transition, $gf\epsilon$. Thus, in absolute abundance analyses of
stars the accuracy of former is never better than the latter. In general, it is
more difficult to measure accurate oscillator strengths for weak transitions
\citep{2003PhST..105...61N}; this is manifested in the fact that a typical error
in experimental $f$-values of weak lines is of the order of $20\%$ or more.
Often one has to rely on oscillator strengths from different sources, affected
by various sources of uncertainties.

In analyses of stars other than the Sun, there is a way to avoid $gf$-values
completely. In the relative abundance analyses, oscillator strengths do not
enter calculations at all, because any abundance estimate derived from a single
line in a spectrum of a metal-poor star is referred to that from same line in a
spectrum of a reference star, e.g. the Sun. However, this works well on
condition that for both objects a line appears on the same part of the
curve-of-growth and, thus, is equally sensitive to various atmospheric
quantities (abundance, pressure and turbulent broadening). This is true only for
a very limited range of stellar parameters.

\begin{figure}
\begin{center}
\hbox{
\resizebox{0.53\columnwidth}{!}{\includegraphics[scale=1]{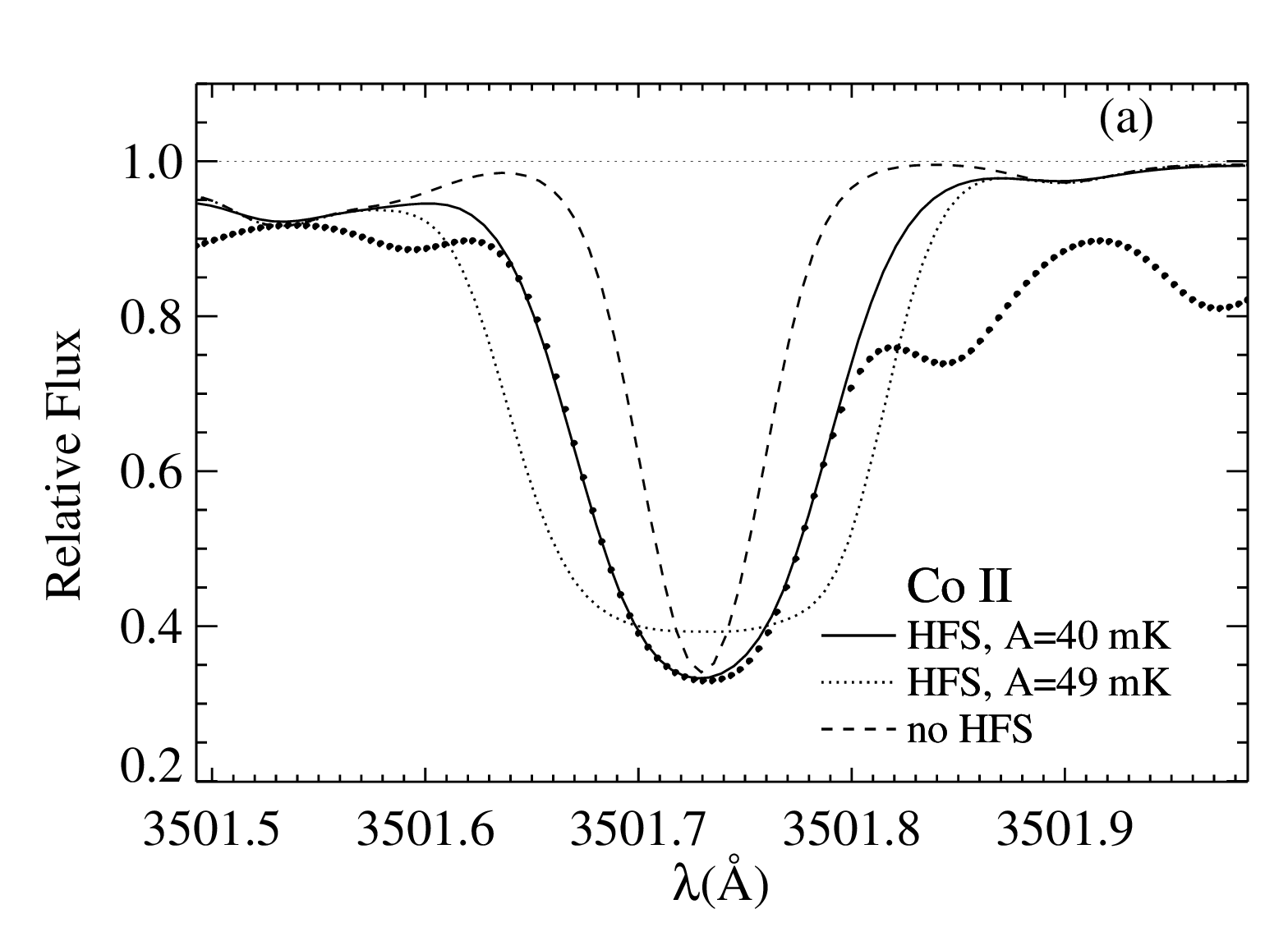}}\hfill
\resizebox{0.53\columnwidth}{!}{\includegraphics[scale=1]{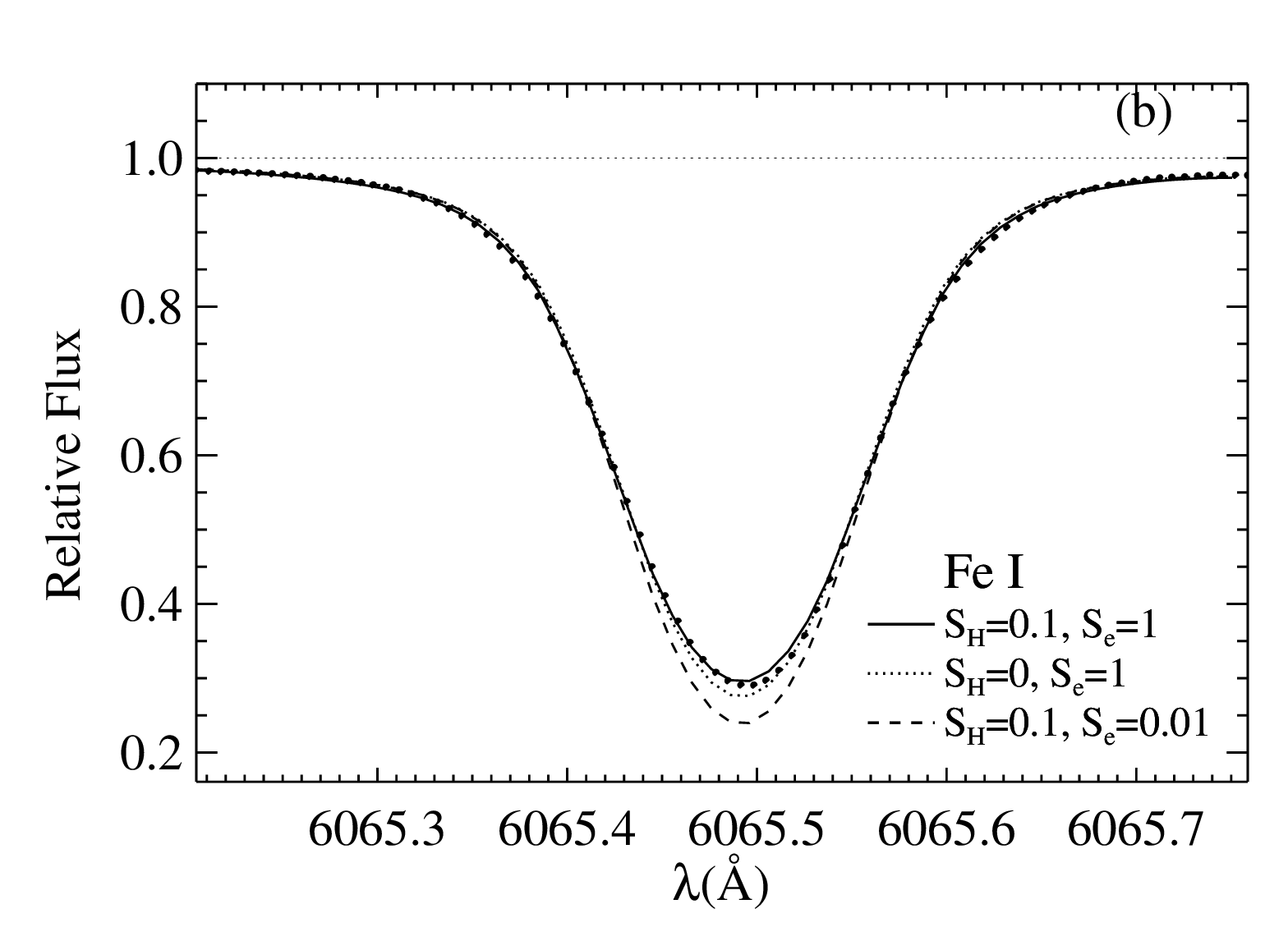}}
}
\caption{(a) The line of \ion{Co}{ii} in the solar KPNO flux spectrum (filled
circles). The profile computed with A $= 49$ mK for the lower level
\Co{a}{5}{P}{}{3} gives an excellent fit of the observed profile (black trace).
The profiles computed neglecting HFS (dashed trace) or with the less accurate
laboratory value A $= 40$ mK (dotted trace) are also shown. (b) The solar line
of \ion{Fe}{i} computed with various scaling factors to inelastic \ion{H}{i} and
e$^-$ collisions, $S_{\rm H}$ respectively $S_{\rm e}$. The observed profile is
shown with filled circles.} 
\label{co_hfs}
\end{center}
\end{figure}

Some lines are affected by hyperfine splitting (HFS) and/or isotopic shift, and
these types of line ``broadening'' are comparable or even larger than the
thermal broadening of lines at local kinetic temperatures in the atmospheres of
solar-type stars. HFS is particularly pronounced for odd-$Z$ elements Mn and Co
with large atomic masses and large nuclear spins, $5/2$ respectively $7/2$. The
effect of HFS is to de-saturate spectral lines leading to a compound profile,
where the strength of each component is linearly proportional to the element
abundance. To compute accurate separation of HFS components, magnetic dipole
constants $A$ must be known with an accuracy of at least $10 \%$. Fig.
\ref{co_hfs} shows that the \ion{Co}{ii} line at $3501$ \AA\ computed with an
approximate estimate $A= 49$ mK for its lower level \Co{a}{5}{P}{}{3}
\citep{1998ApJS..117..261P} cannot fit the observed profile at all. In contrast,
a more accurate laboratory value $A = 40$ mK from \citet{2010MNRAS.401.1334B}
provides an excellent fit to the observed profile. A profile computed without
HFS is also shown.

Finally, accurate wavelengths of transitions are important in studies of line
asymmetries \citep{1981A&A....96..345D,1990A&A...228..155N}, e.g. in diagnostic
analyses by means of 3D hydrodynamical model atmospheres and for identification
of blending features.

\subsection{NLTE radiative transfer}

Computing NLTE line formation is very time-consuming and complex compared to
LTE, because much more atomic data are necessary in addition to the standard set
of parameters for individual lines as described above.

Atomic models for statistical equilibrium calculations must be fairly complete.
Various processes representing interaction between atoms, electrons, and
radiation are to be included. These are photon absorption in line transitions,
photoionization, excitation and ionization by collisions with free electrons and
neutral hydrogen atoms. All processes include their reverse reactions; for
scattering in discrete radiative transitions, one usually assumes complete
frequency redistribution. The current status of NLTE modelling for neutral atoms
of transition metals is described below. These elements have a partly filled
$3d$ subshell that gives rise to fairly complicated but interesting atomic
properties.
\begin{figure}[!ht]
\begin{center}
\resizebox{0.7\columnwidth}{!}
{\includegraphics[scale=1, angle=90]{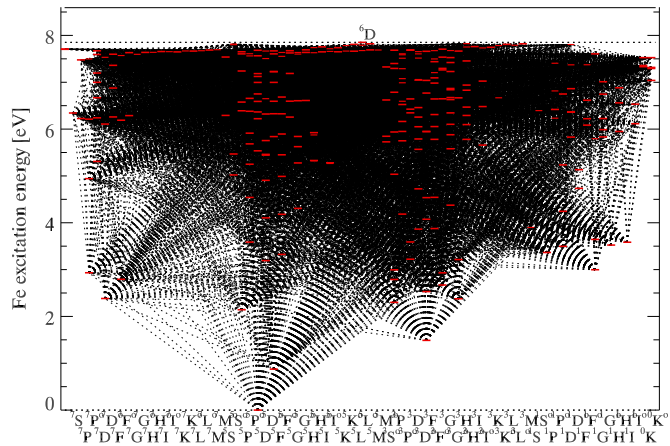}}
\caption{Grotrian diagram of the \ion{Fe}{i} atom constructed using the new
predicted levels and transitions from the Kurucz's database. The ground state of
the \ion{Fe}{ii}, \Fe{a}{6}{D}{}{}, is indicated.}
\label{fe_grot}
\end{center}
\end{figure}

The number of levels and discrete radiative transitions in neutral
atoms of the Fe-group is enormous. Recent calculations of Kurucz predict tens of
thousands of levels and few millions of transitions. Such atomic models
are not tractable even with 1D NLTE codes; thus, recently, we developed
efficient algorithms to combine atomic levels and transitions into super-levels
and super-lines in a way that a reduced model atom inherits the basic
properties of the complete model and has similar performance under restriction
of different stellar parameters. For example, application of these algorithms to
\ion{Fe}{i} reduced the number of levels from $37\,500$ to $292$, and the number
of radiative bound-bound transitions from $\sim 6\,029\,000$ to $13600$ (Fig.
\ref{fe_grot}). Furthermore, for 3D NLTE calculations, we have constructed a
model with $172$ levels and $1120$ transitions, which is now being tested in 1D
and 3D NLTE line formation codes. The requirements to the accuracy of transition
probabilities, or $f$-values, are not very strict. Strong lines, which are
subject to non-equilibrium excitation effects, have laboratory $f$-values with
$5 - 10\%$ accuracy. For transitions between a bulk of intermediate- and
highly-excited levels, the accuracy of $20 - 30 \%$ is sufficient, because the
energy separation of the levels is in most cases small enough for collisions to
dominate over radiative rates.

The situation is more uncertain for bound-free radiative processes. Hydrogenic
approximation is usually adopted for photoionization; however comparison with
quantum-mechanical data for \ion{Fe}{i} \citep{1997A&AS..122..167B} and
\ion{Cr}{i} \citep{2009JQSRT.110.2148N} shows that the former is not correct
even within an order of magnitude. The quantum-mechanical cross-sections for
the levels of both atoms are larger in the background and are characterised by
prominent resonances that leads to significantly enhanced ionization rates in
statistical equilibrium calculations and, thus, to different abundances.
In some cases, accurate position of resonances is important, because UV
radiation field varies strongly with frequency in the atmospheres of late-type
stars and ionization rates may change depending on whether a resonance appears
at energies with high or low fluxes. Available laboratory data very sparse; e.g.
for a few Ti I intermediate-excitation levels the cross-sections were measured
by \citet{2009ChJCP..22..615Y} by the method of resonance ionization mass
spectrometry.
Experimental values are very useful to check the accuracy of theoretical values
for individual transitions, however they cannot cover the complete atomic system
that is essential in NLTE calculations of atomic level populations for stellar
atmosphere studies. Thus, quantum-mechanical data are indispensable.

Cross-sections for \ion{H}{i} and e$^-$ collisions are basically unknown. Thus,
we have to rely on commonly-used approximations or to neglect these processes 
completely. Cross-sections for continuum and allowed discrete transitions due to
inelastic collisions with \ion{H}{i} are computed with the formulas of
\citet{1968ZPhy..211..404D,1969ZPhy..225..470D,1969ZPhy..225..483D}. This recipe
was originally developed for collisions between equal atoms, H and Ar, and later
it was extended to Li-H collisions by \citet{1984A&A...130..319S}. The rates of
allowed and forbidden transitions due to collisions with electrons are
calculated from the formulae of \citet{1962ApJ...136..906V} and
\citet{1973asqu.book.....A}, respectively. Bound-free transitions caused by
electron collisions are treated according to \citet{1962amp..conf..375S}.
Similar to photoionization, the accuracy of the data obtained with these
formulae is very low. Comparison of R-matrix calculations for e$^-$ collisions
for Ca II \citep{2007A&A...469.1203M} with the formula of
\citet{1962ApJ...136..906V} reveals order of magnitude differences. What
concerns inelastic \ion{H}{i} collision, compared to the Drawin's formulae, ab
initio quantum-mechanical calculations predict significantly lower collision
rates for certain transitions of simple alkali atoms
\citep{2003PhRvA..68f2703B,2010A&A...519A..20B}, and they show that, in
addition to excitation, other effects like ion-pair formation become important.
Excitation and ionization balance in the atoms of the Fe-group depend on e$^-$
collision efficiency for solar-type stars, i.e. metallicities [Fe/H] $\geq
-0.5$. In the atmospheres of metal-poor stars, where NLTE effects on abundances
are large, collisional excitation is fully controlled by neutral hydrogen.

The deficiencies in NLTE models often reveal themselves, when synthetic spectra
are compared with observations, e.g. by producing large abundance discrepancies
between different spectral lines of the same atom. Thus, it is common to
introduce scaling factors, which are adjusted to obtain agreement between
various spectral lines. Fig. \ref{co_hfs}b shows the profiles of the solar
\ion{Fe}{i} line computed with different scaling factors to Drawin's \ion{H}{i}
collision rates, $S_{\rm H} = 0, 0.1$. Inelastic $e^-$ collision rates computed
using the formulae of \citet{1962ApJ...136..906V} and
\citet{1962amp..conf..375S} are also scaled by $S_{\rm e} = 0.01, 1$. The figure
shows that a proper choice of $S_{\rm H}$ and $S_{\rm e}$ may also remove
discrepancy between observed and synthetic line profiles.

\section{Discussion and outlook}

The calculations of NLTE stellar spectra for the late-type stars are usually
decoupled from calculations of model atmospheres. As a rule, the latter are
computed assuming LTE, and line formation calculations are performed keeping the
input model fixed. Such a ``simplified'' NLTE approach is not fully justified,
because various elements affect the structure of the atmosphere by contributing
free electrons and/or opacity, such as Mg, Al, Si, Fe, Ti (see Sec.
\ref{sec:opac}). Deviations from LTE in the excitation and ionization equilibria
of these elements will lead to changes in the energy balance. Thus, NLTE rate
equations for all these elements must be solved simultaneously in model
atmosphere calculations, a numerically challenging task. Another problem is that
atomic data for some of these elements are of very low quality or lack
completely. This refers to photoionization, \ion{H}{i} and e$^-$ impact
excitation and ionization cross-sections.

The first problem has been successfully overcome with enormous progress in
numerical techniques and computing facilities in the past $10$ years. At present
it is possible to compute NLTE line-blanketed model atmospheres
\citep{1989ApJ...339..558A,2005ApJ...618..926S,2008A&A...492..833H}. These and
other studies proved the importance of NLTE effects on the atmospheric structure
and spectral energy distributions for the Sun \citep{2005ApJ...618..926S} and
non-negligible differences for metal-poor stars \citep{2005ESASP.560..967S}.
There is no doubt that soon self-consistent NLTE models will replace LTE-based
ones; accurate atomic data will stimulate and, thus, accelerate this transition.

\bibliographystyle{asp2010}
\bibliography{references}

\begin{thebibliography}{}
\expandafter\ifx\csname natexlab\endcsname\relax\def\natexlab#1{#1}\fi
\expandafter\ifx\csname url\endcsname\relax
  \def\url#1{\texttt{#1}}\fi
\expandafter\ifx\csname urlprefix\endcsname\relax\def\urlprefix{URL }\fi
\providecommand{\eprint}[2][]{\url{#2}}

\bibitem[{{Allen}(1973)}]{1973asqu.book.....A}
{Allen}, C.~W. 1973, {Astrophysical quantities} (London: University of London,
  Athlone Press, 1973, 3rd ed.)

\bibitem[{{Anderson}(1989)}]{1989ApJ...339..558A}
{Anderson}, L.~S. 1989, \apj, 339, 558

\bibitem[{{Barklem} et~al.(2010){Barklem}, {Belyaev}, {Dickinson}, \&
  {Gad{\'e}a}}]{2010A&A...519A..20B}
{Barklem}, P.~S., {Belyaev}, A.~K., {Dickinson}, A.~S., \& {Gad{\'e}a}, F.~X.
  2010, \aap, 519, A20+. \eprint{1006.5164}

\bibitem[{{Bautista}(1997)}]{1997A&AS..122..167B}
{Bautista}, M.~A. 1997, \aaps, 122, 167

\bibitem[{{Bell} et~al.(2001){Bell}, {Balachandran}, \&
  {Bautista}}]{2001ApJ...546L..65B}
{Bell}, R.~A., {Balachandran}, S.~C., \& {Bautista}, M. 2001, \apjl, 546, L65

\bibitem[{{Belyaev} \& {Barklem}(2003)}]{2003PhRvA..68f2703B}
{Belyaev}, A.~K., \& {Barklem}, P.~S. 2003, \pra, 68

\bibitem[{{Bergemann} et~al.(2010){Bergemann}, {Pickering}, \&
  {Gehren}}]{2010MNRAS.401.1334B}
{Bergemann}, M., {Pickering}, J.~C., \& {Gehren}, T. 2010, \mnras, 401, 1334.
  \eprint{0909.2178}

\bibitem[{{Dravins} et~al.(1981){Dravins}, {Lindegren}, \&
  {Nordlund}}]{1981A&A....96..345D}
{Dravins}, D., {Lindegren}, L., \& {Nordlund}, A. 1981, \aap, 96, 345

\bibitem[{{Drawin}(1968)}]{1968ZPhy..211..404D}
{Drawin}, H.-W. 1968, Zeitschrift fur Physik, 211, 404

\bibitem[{{Drawin}(1969{\natexlab{a}})}]{1969ZPhy..225..470D}
{Drawin}, H.~W. 1969{\natexlab{a}}, Zeitschrift fur Physik, 225, 470

\bibitem[{{Drawin}(1969{\natexlab{b}})}]{1969ZPhy..225..483D}
--- 1969{\natexlab{b}}, Zeitschrift fur Physik, 225, 483

\bibitem[{{Grupp}(2004)}]{2004A&A...420..289G}
{Grupp}, F. 2004, \aap, 420, 289

\bibitem[{{Grupp} et~al.(2009){Grupp}, {Kurucz}, \&
  {Tan}}]{2009A&A...503..177G}
{Grupp}, F., {Kurucz}, R.~L., \& {Tan}, K. 2009, \aap, 503, 177.
  \eprint{0906.5449}

\bibitem[{{Haberreiter} et~al.(2008){Haberreiter}, {Schmutz}, \&
  {Hubeny}}]{2008A&A...492..833H}
{Haberreiter}, M., {Schmutz}, W., \& {Hubeny}, I. 2008, \aap, 492, 833.
  \eprint{0810.3471}

\bibitem[{{Johansson}(2008)}]{2008psa..conf..123J}
{Johansson}, S. 2008, in Precision Spectroscopy in Astrophysics, edited by
  {N.~C.~Santos, L.~Pasquini, A.~C.~M.~Correia, \& M.~Romaniello }, 123

\bibitem[{{Mashonkina}(2009)}]{2009PhST..134a4004M}
{Mashonkina}, L. 2009, Physica Scripta Volume T, 134, 014004

\bibitem[{{Mel{\'e}ndez} et~al.(2007){Mel{\'e}ndez}, {Bautista}, \&
  {Badnell}}]{2007A&A...469.1203M}
{Mel{\'e}ndez}, M., {Bautista}, M.~A., \& {Badnell}, N.~R. 2007, \aap, 469,
  1203. \eprint{0704.3807}

\bibitem[{{Nahar}(2009)}]{2009JQSRT.110.2148N}
{Nahar}, S.~N. 2009, JQSRT, 110, 2148

\bibitem[{{Nilsson} et~al.(2003){Nilsson}, {Ivarsson}, {Sabel}, {Sikstr{\"o}m},
  \& {Curtis}}]{2003PhST..105...61N}
{Nilsson}, H., {Ivarsson}, S., {Sabel}, H., {Sikstr{\"o}m}, C.~M., \& {Curtis},
  L.~J. 2003, Physica Scripta Volume T, 105, 61

\bibitem[{{Nordlund} \& {Dravins}(1990)}]{1990A&A...228..155N}
{Nordlund}, A., \& {Dravins}, D. 1990, \aap, 228, 155

\bibitem[{{Pickering} et~al.(1998){Pickering}, {Raassen}, {Uylings}, \&
  {Johansson}}]{1998ApJS..117..261P}
{Pickering}, J.~C., {Raassen}, A.~J.~J., {Uylings}, P.~H.~M., \& {Johansson},
  S. 1998, \apjs, 117, 261

\bibitem[{{Seaton}(1962)}]{1962amp..conf..375S}
{Seaton}, M.~J. 1962, in Atomic and Molecular Processes, edited by D.~R.
  {Bates}, 375

\bibitem[{{Short} \& {Hauschildt}(2005{\natexlab{a}})}]{2005ApJ...618..926S}
{Short}, C.~I., \& {Hauschildt}, P.~H. 2005{\natexlab{a}}, \apj, 618, 926.
  \eprint{arXiv:astro-ph/0409693}

\bibitem[{{Short} \& {Hauschildt}(2005{\natexlab{b}})}]{2005ESASP.560..967S}
--- 2005{\natexlab{b}}, in 13th Cambridge Workshop on Cool Stars, Stellar
  Systems and the Sun, edited by {F.~Favata, G.~A.~J.~Hussain, \& B.~Battrick},
  vol. 560 of ESA Special Publication, 967

\bibitem[{{Steenbock} \& {Holweger}(1984)}]{1984A&A...130..319S}
{Steenbock}, W., \& {Holweger}, H. 1984, \aap, 130, 319

\bibitem[{{van Regemorter}(1962)}]{1962ApJ...136..906V}
{van Regemorter}, H. 1962, \apj, 136, 906

\bibitem[{{Yang} et~al.(2009){Yang}, {Hu}, {Wu}, {Fan}, {Cong}, {Cheng}, {Ji},
  {Yao}, {Zheng}, \& {Cui}}]{2009ChJCP..22..615Y}
{Yang}, J., {Hu}, X., {Wu}, H., {Fan}, J., {Cong}, R., {Cheng}, Y., {Ji}, X.,
  {Yao}, G., {Zheng}, X., \& {Cui}, Z. 2009, Chinese Journal of Chemical
  Physics, 22, 615

\end{thebibliography}

\end{document}